# PROPELLING INTERPLANETARY SPACECRAFT UTILIZING WATER-STEAM

Jorge Martinez,* and Jekan Thangavelautham†

Beyond space exploration, there are plans afoot to identify pathways to enable a space economy, where human live and work in space. One critical question is what are the resources required to sustain a space economy? Water has been identified as a critical resource both to sustain human-life but also for use in propulsion, attitude-control, power, thermal and radiation protection systems. Water may be obtained off-world through In-Situ Resource Utilization (ISRU) in the course of human or robotic space exploration that replace materials that would otherwise be shipped from Earth." Water has been highlighted by many in the space community as a credible solution for affordable/sustainable exploration. Water can be extracted from the Moon, C-class Near Earth Objects (NEOs), surface of Mars and Martian Moons Phobos and Deimos and from the surface of icy, rugged terrains of Ocean Worlds. However, use of water for propulsion faces some important technological barriers. A technique to use water as a propellant is to electrolyze it into hydrogen and oxygen that is then pulse-detonated. High-efficiency electrolysis requires use of platinum-catalyst based fuel cells. Even trace elements of sulfur and carbon monoxide found on planetary bodies can poison these cells making them unusable. In this work, we develop steam-based propulsion that avoids the technological barriers of electrolyzing impure water as propellant. Using a solar concentrator, heat is used to extract the water which is then condensed as a liquid and stored. Steam is then formed using the solar thermal reflectors to concentrate the light into a nanoparticle-water mix. This solar thermal heating (STH) process converts 80 to 99% of the incoming light into heat. In theory, water can be heated to 1000 K to 3000K with a resulting Isp from 190s to 320s. This propulsion system can offer higher thrust than current electrical propulsion methods and represents a high delta-v solution for small spacecrafts. A further understanding of the concentration system, implications for GNC and the heat transfer process in the nanofluid is presented in this work.

## INTRODUCTION

The space industry is approaching a critical time of diversification away from mainly military presence and science-driven exploration into commercial ventures including space tourism, space resource prospecting/mining and space services. This transformation into a full-fledged space economy will result in full range of activities and the use of resources that create value and benefits to human beings, in the course of exploring, researching, understanding, managing, and utiliz-

* Masters Candidate, Aerospace and Mechanical Engineering, Univ. of Arizona, 1130 N Mountain Ave., Tucson.
†† Assistant Professor, Aerospace and Mechanical Engineering, Univ. of Arizona, 1130 N Mountain Ave., Tucson.

ing space. If this venture is to succeed, humanity must figure out a solution for an affordable/sustainable exploration.

Two main factors have been identified as critical to advancing space technology. First, the miniaturization of computer electronics, sensors, guidance, navigations and control devices, and the proliferation of low-cost, high-efficiency power systems on space have led to the development of ever smaller spacecrafts, drastically reducing launch and development costs. The second critical factor is the development of space systems and architecture that can partially or fully exploit space resources to lower cost of operation and enable sustainability. Water has been highlighted by many in the space community as a critical resource (Figure 1), not only to sustain human-life but also for igniting the space economy because of its multifunctional character. It has applications in propulsion, power, thermal storage and radiation protection systems. Furthermore, water may be obtained off-world through In-Situ Resource Utilization (ISRU). Water can be extracted from the Moon, C-class Near Earth Objects (NEOs), surface of Mars and Martian

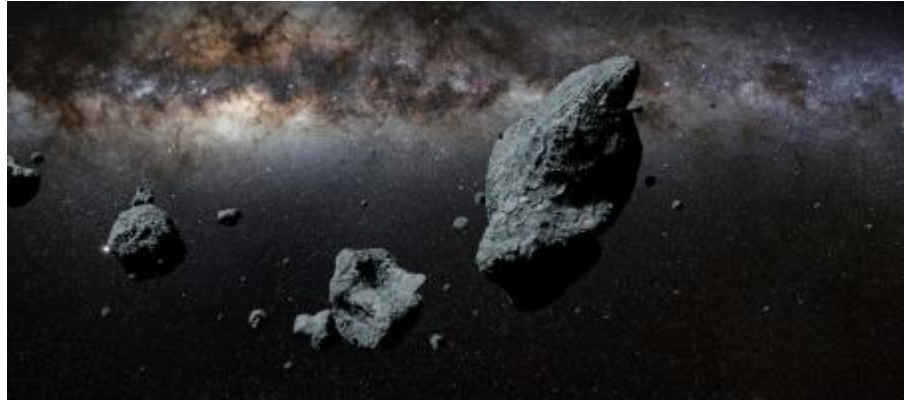

**Figure 1:** The asteroid, Moon and Mars all hold significant quantities of water that can kick-start a space economy.

Moons Phobos and Deimos and from the surface of icy, rugged terrains of Ocean Worlds.

Water is a compelling choice for rocket fuel, as it can be easily stored and is green and non-volatile. This propellant might be used for both interplanetary travel and surface exploration in a mother-daughter architecture concept. Interplanetary missions are expected to be significantly cheaper and routine if there are strategically located refueling stations. Mission to and from Mars can readily exploit this concept. Further ground robots that require propulsion to traverse extreme terrain can use this technology [21,22].

However, use of water for propulsion faces some important technological barriers. Water electrolyzed into hydrogen and oxygen would appear to provide a high-performance propulsion system. There are indeed challenges in storing the hydrogen and oxygen in liquid form over prolonged periods and over varying environmental temperatures. This is especially difficult for miniature spacecraft where there is limited mass and volume for cryogenic storage. Another option is to electrolyze the water on demand utilizing electrolysis[20]. The hydrogen and oxygen is pulse-detonated. High-efficiency electrolysis requires use of platinum-catalyst based fuel cells. Even trace amounts of sulfur, carbon-monoxide and other impurities thought to be found on C-type asteroids can poison platinum catalyst sites irreversibly. Over the past 20 years, solutions to protecting platinum catalysts from chemical poisoning has been met with very little success. The alternative is to purify the water using reverse-osmosis or steam-condensation. The net result is significant expenditure of energy. Reverse osmosis which is overall more energy efficient of the two options typically requires 3-4 Wh/kg brine water. A credible alternative is to use superheated water-steam as propulsion. This system is not impacted by water impurities: water can be extracted as ice, heated and stored as liquid, and then employed as superheated steam for propulsion. Another alternative is super-heated hydrogen which although is more difficult to store and obtain makes up for significantly increased Isp. This paper builds upon our earlier paper on solar thermal propulsion and gets into the details of the propulsion system design [17].

As noted earlier, use of small spacecrafts for interplanetary travel could drastically reduce mission costs. However, small spacecraft missions, are largely confined to Low Earth Orbit (LEO), due to the mass restriction of high-efficiency propulsions systems. Although some high delta-v solutions have been developed for these spacecrafts, the limited thrust entails longer waiting times and precise maneuvers for interplanetary missions. However, a high-thrust, high-impulse propulsion technology, enabled by off-world resources and access to strategic refueling depots would allow small spacecraft to be the work-horse (tugboats) for interplanetary travel.

In the following sections, we present background and related work on solar thermal propulsion updated description of the solar thermal propulsion system for a small spacecraft followed by analysis, discussion, conclusions and future work.

**BACKGROUND**

Solar concentrators have been widely studied as means to collect energy on Earth. In this section, the mechanism and types of solar concentration technologies are shown. Radiation is emitted constantly by all substances as a result of the molecular and atomic agitation associated with the internal energy of the material[1]. At an equilibrium state, this internal energy is proportional to the temperature of the substance. One of the main factors for the importance of thermal radiation is the way in which radiant emission depends on temperature. For conduction and convection, the heat transfer between two locations depends linearly on the temperature. However, the transfer of thermal radiation energy depends on the forth power of the temperature difference between the two points[1]. It is for this difference that radiation becomes the most important way of heat transfer at high temperature levels. Besides, radiation transfer can occur without any medium between the two locations, in contrast to conduction and convection. This means that, when no medium is present, radiation becomes the only significant mechanism of heat transfer.

Solar energy is the most abundant energy resource on earth, and the solar irradiance reaching the earth's surface can be up to 1 kW/m$^2$ under ideal conditions. Many applications require energy at higher temperatures than those reached from incident solar radiation. For that, solar energy is concentrated in collectors that capture and focus the solar radiation onto a smaller receiving surface, achieving higher working temperatures.

There is a variety of concentrating solar thermal technologies available. Combining different concentrators and receiver, four main technologies can be found to concentrate solar power: parabolic trough (PT), linear Fresnel reflector (LT), central receiver or solar tower (ST) and parabolic dish (PD) (Figure 2). They are distinguished by the way they focus the sun's rays and the technology used to receive the solar energy. Usually, they can be classified according to the focus type (linear focus or point focus), the receiver type (fixed or mobile), or the concentration level (medium or high). In solar tower and linear Fresnel technologies, the receiver remains fixed and mechanically independent from the concentration system. On the contrary, both in parabolic trough and parabolic dish, the receiver and concentration system move together, enabling optimal arrangement, and leading to an increased optical efficiency.

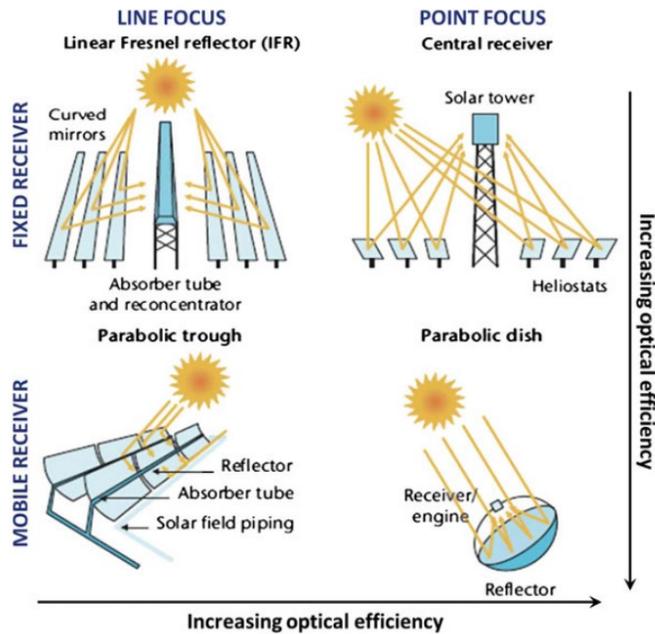

**Figure 2:** Efficiency of Earth solar concentrating technologies[18].

These same principles could be applied in space, to capture sunlight and convert it into heat. Around Earth, solar irradiance is 1365 W/m$^2$, while near Mars it is only 590 W/m$^2$. In space, there are ways to modify the concentrator systems to make them scale well- namely using inflatable systems[2]. These systems have minimal launch footprint in terms of mass and volume. Although inflatable Fresnel lens have been used with success in space[3] their concentration ratio isn't as attractive as membrane-type inflatable concentrators.

**RELATED WORK**

Solar thermal propulsion (STP) has been under development since the 1950s, when it was first proposed as a solution to eliminate the oxidizer, and therefore reduce launch mass. These systems usually need to storage cryogenic propellants, carry important power losses in eclipse time, and need a separate thermal collection system in addition to photovoltaics or batteries.

A solar thermal rocket carries the means of capturing solar energy such as concentrators and mirrors. In the case of photovoltaic systems, solar energy is converted to electrical power. In here, instead the solar energy is directly used as heat. The heated propellant is fed through a conventional rocket nozzle to produce thrust. Typically, a low molecular weight propellant is used, such as hydrogen, corresponding to a high specific impulse. Solar thermal propulsion effectively bridges the performance between the chemical and electrical propulsion. As stated in the introduction, solar thermal propulsion represents the performance between chemical and state-of-the-art solutions for small spacecrafts.

Therefore, early solar thermal propulsion designs were focused on large spacecrafts that presumed that solar thermal energy would be only utilized for propulsion means. K. Ehricke[4] proposed the first STP concept in 1956. The proposed concept consisted only on the three primary components for any solar heating mechanism: the solar concentrator, a receiver/absorber, and the propellant mechanism. In that concept, the spacecraft gross weight was around 7000 kg, with two 39m diameter inflatable spherical solar concentrators, and a liquid hydrogen propellant mass fraction of 70% that could only be heated up to 1000 K. The architectures resulting from this development ended up being highly complex and prevented an in-space demonstration.

These STP challenges were tackled for first time in 1963, when Electro-Optical Systems (EOS), conducted the first detailed investigation of a smaller-scale technology. Experimentally, they successfully proved that hydrogen temperatures around 2300 K could be achieved with a solar concentrator of only 1.5 m, using a solar heated molybdenum absorber and tungsten flow tube[5]. Rocketdyne International, contracted by the Air Force Rocket Propulsion Laboratory (AFRPL) produce a prototype receiver and thruster, that became the foundation for modern solar thermal research efforts. The final design, with a 25 × 25 ft solar furnace, was tested and achieved hydrogen temperatures of 1810 K, corresponding to an estimated Isp of 650s[6]. Since then, multiple solar rocket technology programs with funding from the Air Force, NASA and other agencies have developed a large technical database.

In 1979 the Air Force sponsored a program to develop an off-axis inflated concentrator solar thermal propulsion system. In this design configuration, the concentrator and the receiver are optically coupled, with the absorber located at the concentrator focus. This direct solar thermal propulsion involves heating the propellant directly.

In 1992, Venkateswaran et al.[7] developed a direct solar collection system combined with a re-radiator wall that captured secondary radiation from the working fluid. They predicted average temperatures ranges from 2200K to 2500K with power levels up to 10 MW. However, the complexity and size of the system made scientific community abandon this idea. In the late 90s, Alexander et al.[8] presented a direct gain solar thermal upper stage engine that heated hydrogen using up to 10kW of concentrated energy. For a nominal hydrogen flow rate of 2lb/hr, they found temperatures around 2500K, claiming they could be raised to 2800K cutting the flow rate to the half. Boeing and AFRL during the 2000s developed Solar Orbit Transfer Vehicle Concept (Figure 3), a low-cost approach to boost heavy vehicles into higher orbits[19].

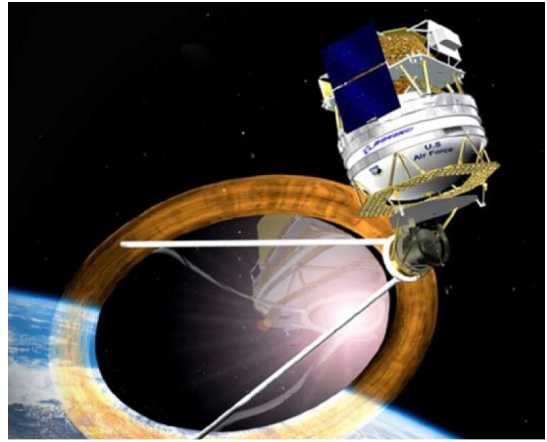

**Figure 3:** Boeing-AFRL Solar Orbit Transfer Vehicle concept that utilizes solar-thermal energy to raise the orbit of large satellites.[19]

In 1996, selective absorptive coating came into the picture as means to increase the solar absorption efficiency in the receiver. In Fortini et al.[9] the sunlight was focused into a cylindrical blackbody cavity, where the inner walls would absorb the heat. On the other side of the inner wall, a high surface area open-cell foam acted as a heat exchanger to transfer the heat from the wall to the hydrogen. Following this trend, selective absorbers were developed on Earth for high-efficiency solar thermal conversion, employing spectrally selective solar absorbers that exhibit a near-blackbody absorption in the solar spectrum while suppressing infrared emission at elevated temperatures. However, developing cost-effective and large-scale solar selective absorbers able to work at high-temperatures and with high efficiency remains a challenge.

Researchers in this field[10] demonstrated that metal-based wafer-scale nanophotonic solar selective absorbers, using a template stripping method, can drastically increase the output energy and decrease fabrication cost. Employing a combination of nickel nanopyramid structures and aluminum oxide coating, maximum absorbance was achieved near the 1-micron spectrum. The theoretical solar-to-electric energy conversion efficiency was observed to be higher than 68%, at the solar concentration of 1000, and working temperatures up to $700°C$. For working temperatures above $700°C$ Shimizu et al.[11] introduced a transparent conductive oxide (TCO) film coated metal microstructures as solar selective absorbers.

An alternative to these direct absorption methods is storage of the heat in thermal storage device. In 2011 Schafer et al.[12] suggested a thermal storage system combined with a means of thermal-to-electric conversion, with the idea of providing a dual-mode power and propulsion system based on thermal energy. Using a spherical concentrator and a Heliostat, boron was used, achieving local temperatures higher than 2570 K. More recently, Nakamura et al.[13] proposed an indirect heating technology, using fiber optics to transfer the light from the concentrator to the solar-thermal convertor, with power transfer up to 200 W and receiver temperatures over 1400 K.

**SOLAR THERMAL PROPULSION**

First and foremost, a high-performance solar concentration is needed in any STP system. Solar concentration technology has been deeply studied for terrestrial solar energy purposes since ancient times. Although shorter, the development of solar concentrator for Solar Thermal Propulsion has also a long history. In this history, the vast majority of concentrator developments have been focused on inflatable structures for large satellites, the first model proposed having a 39m diameter. The only rigid structures, proposed by the ISUS program, had densities of 2.5 $kg/m^2$. With demonstrated area densities below $1\ kg/m^2$ for radio transmission inflatables structures, this technology allows promising packaging capabilities. Almost 15 years ago, Sahara et al.[14] in collaboration with JAXA, developed a lightweight solar concentrator that seems the most promising solution to the date. In this work, they created a thin film polymer and polyamide solar concentrator with areas densities of only $180\ g/m^2$. A high temperature vacuum process was used to form them onto paraboloidal glass molds. Forming error was also taken into account, and the resulting structure was allowed to set for several days minimizing stress buildup. As a result, high efficiency reflectors with concentration ratios over 10,000:1 were achieved. Along with this technology, a solar thermal engine out of single crystal refractory metals was combined into a concept for a de-orbiting module[15].

We believe this technology represents the future of solar concentrators. Apart from having a superior concentrating performance than a rigid structure, a microsatellite based solar platform would require less than $2\ m^2$ of total concentrator area, and therefore the use of self-supporting thin structures is possible. However, a rigid concentrator might represent a simpler and more compact solution, as it does not need supporting structures. Experiments[16] in lower solar concentrations have shown that up to 20 ml of a gold nanofluid solution can be converted into steam in 5 minutes, under 280 light flux concentrated into a 28 $cm^2$ focal area. According to this experiment, the steam generation at a given particles concentration is proportional to the concentrated solar flux, and the focal area. Increasing the focal area, for a given concentration ratio, means increasing the parabolic area.

Based on the calculations of the previous subsection, Figure 4 (left) shows a solar concentrator concept. In this case, the desired area was $1m^2$ and therefore the radius of the top surface was computed to be $R = 0.56\ m$, and the focal length $f = 0.4\ m$. With these parameters, a concentration ratio of $C = 10.000$ can be achieved. A mirror is then located in the focal area of the concentrator, to reflect the sunlight towards the receiver, located in the satellite. The reason for this is that the size of the hot chamber would be restricted if it must be located in the focus itself. Furthermore, better heat isolation is possible if the chamber containing the heated water is located in the satellite. From the hot chamber, the superheated steam is directly expanded through a nozzle to produce thrust, as pictured in Figure 4 (right).

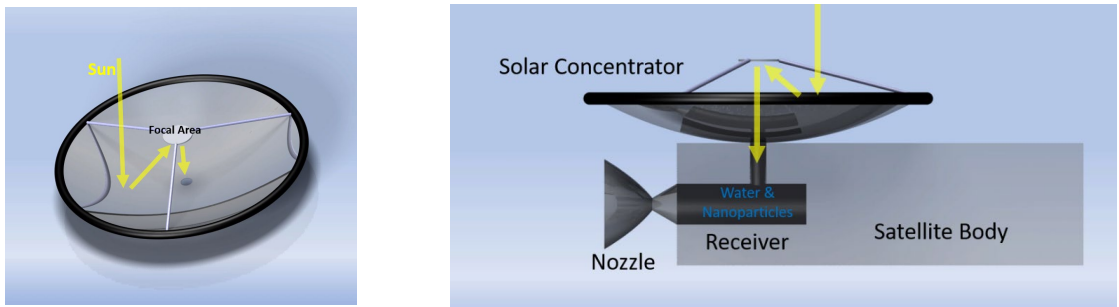

**Figure 4:** (Left) Solar thermal concentrator. (Right) Solar thermal propulsion system concept for a small satellite.

Furthermore, this configuration allows for a possible thermal storage system by introducing a secondary mirror in the concentrated light trajectory. This way, energy could be collected when facing the sun, either to pre-heat the fluid when propulsion is required, or to supply power to the rest of the spacecraft. The concentrated light is absorbed into the receiver, which is the component responsible of converting this solar energy into heat. Traditional solar thermal collectors cap-ture light energy on an absorbing surface. This energy has to then be transferred to a circulating fluid via convection.

Currently, solar thermal collectors capture light energy on an absorbing surface, which must then transfer that energy via conduction or convection to a circulating fluid. In an indirect gain concept, this fluid will transfer the energy to the propellant through a heat exchanger. As shown in Figure 5, this thermal path can be substantially shorter if a nanofluid is employed as the working fluid.

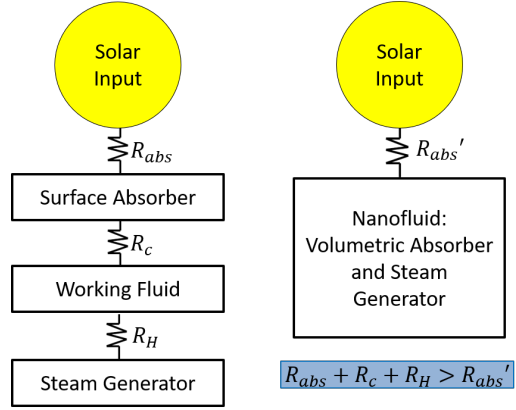

**Figure 5:** Comparison of a simplified thermal resistance model. $R_{abs}$, $R_c$ and $R_H$ correspond to the absorption, conductive/convective and heat exchanger resistances respectively.

**RESULTS**

In this section, the performance and issues that this system may face are discussed. As stated in the previous section, experimental work in high flux intensities and carbon nano-particles are needed for a better understanding of the behavior at high temperatures. In here, we propose an initial sizing and performance based on simplified models

**Solar Collector Performance**

The solar flux incident on the parabolic dish will heat the receiver. This solar heat is dependent on the concentration ratio, and the higher this value, the higher temperature and specific impulse the system will provide. An approximation is still needed to know what working temperatures can we theoretically achieve with the system. Assuming a perfect receiver, the maximum temperature of the carbon nanoparticles can be estimated by assuming that at equilibrium, the concentrated heat flux from the sun should be equal to the radiative losses of the receiver, assuming no losses by convection:

$$T_R^4 = (\frac{\alpha \gamma C I_{solar}}{\varepsilon \sigma} + T_{amb}^4) \qquad (1)$$

where $\alpha$ is the absorptivity of the receiver, $\gamma$ is the reflectivity of the concentrator, $C$ is the concentrator ratio, $I_{solar}$ is the incident solar irradiance, $\varepsilon$ is the receiver emissivity and $\sigma$ is the Stefan Boltzmann constant. The optical properties of the receiver are however a function of the temperature. When the temperature starts rising, the absorber glows, which reduces the efficiency of light absorption. Carbon-based nanoparticles maintain an 80% of absorptivity efficiency at temperatures higher than 2000K[17].

**Table 1. Solar Collector Optical Properties.**

| Carbon-Black Receiver |
|---|

| | | | |
|---|---|---|---|
| Incident Solar Flux | $I_{solar}$ | Receiver Absorptivity | $\alpha = 0.97$ |
| Parabolic Dish Reflectivity | $\gamma = 0.$ | S.B. constant $[W/m^2K^{-4}]$ | $\sigma = 5.67 \cdot 10^{-8}$ |
| Receiver Emissivity | $\varepsilon = 0.$ | Ambient Temperature | $T_{amb} = 2.7\ K$ |

Based on this value, the maximum temperature that a carbon-based receiver could achieve is shown on Figure 6 (left). For that, we need to define the efficiency of energy collection based on the receiver temperature. This one, can be determined as the fraction between the useful energy in the receiver and the incident energy from the concentrator:

$$Q_{useful} = A_p \left[ \alpha \gamma C I_{solar} - \varepsilon \sigma \left( T_R^4 - T_{amb}^4 \right) - U_L (T_R - T_{amb}) \right] \quad (2)$$

$$Q_{in} = A_p \gamma C I_{solar} \quad (3)$$

$$\eta = \frac{Q_{useful}}{Q_{in}} = \alpha - \frac{\varepsilon \sigma \left( T_R^4 - T_{amb}^4 \right)}{\gamma C I_{solar}} - \frac{U_L (T_R - T_{amb})}{\gamma C I_{solar}} \quad (4)$$

As indicated in Figure 6 (right), the collection efficiency decreases with the receiver temperature. This is because radiation losses in the receiver will increase, until the point where the receiver can't be further heated. However note that this efficiency increases for higher concentration ratios.

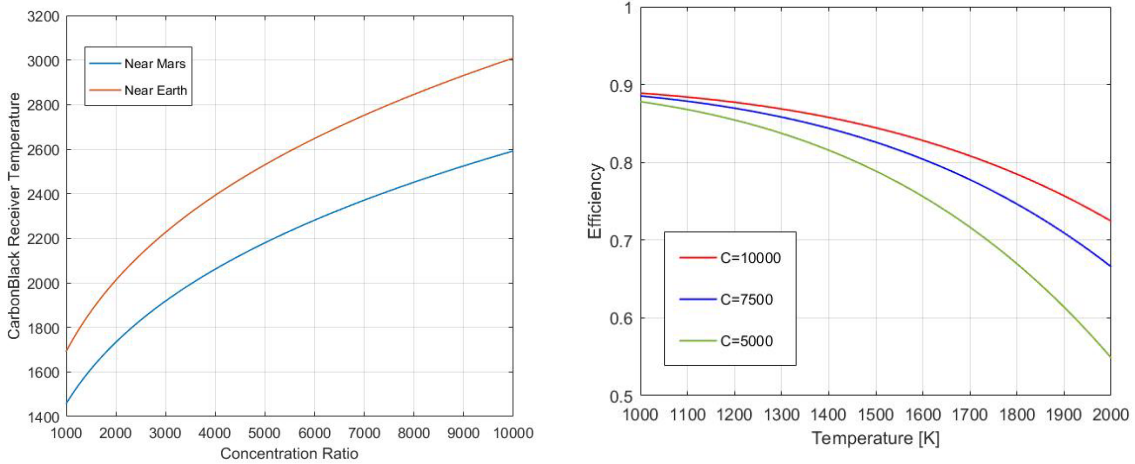

**Figure 6:** (Left) Receiver Temperature vs Concentration Ratio near Mars and Earth. (Right) Collection efficiency vs Receiver temperature.

Theoretically, this system should be able to heat up water steam to almost 3000 K under Earth solar irradiance, and a maximum concentration ratio of 10.000. This high temperature would mean a high specific impulse, as shown in Figure 8 (left). However, these temperatures can only be achieved under low collection efficiencies, as a lot of energy will be lost in terms of radiation. This low absorption efficiency means that the time required to achieve that temperature, for a given propellant mass, can become high. This is would mean low mass flow rates, and therefore lower thrusts. The system design suggests therefore a compromise between high thrust and high specific impulse. This ability to dial up or down the specific impulse and thrust might open interesting performance characteristics.

**Nozzle Performance**

The ANSYS Fluent model was solved in order to compute the real performance of the designed nozzle. Once the solution converges, the velocity and temperature contours can be easily plot. The velocity contour is shown in Figure 7 (left). In these figures it can be seen how the friction due to the boundary layer in the walls makes the velocity step out to zero in the wall (no-slip boundary condition), which will create thrust and impulse losses. The evolution of the Mach number within the nozzle is also shown in Figure 7 (right).

Note that the exhaust velocity is very similar to the one computed analytically (2780 m/s). In the ANSYS model however, the velocity has to be averaged over the exhaust area. ANSYS surface integrals allows us to calculate $I_{sp}$ this way. The computed $I_{sp}$ this way is 267 m/s, only a 5% lower than the analytical.

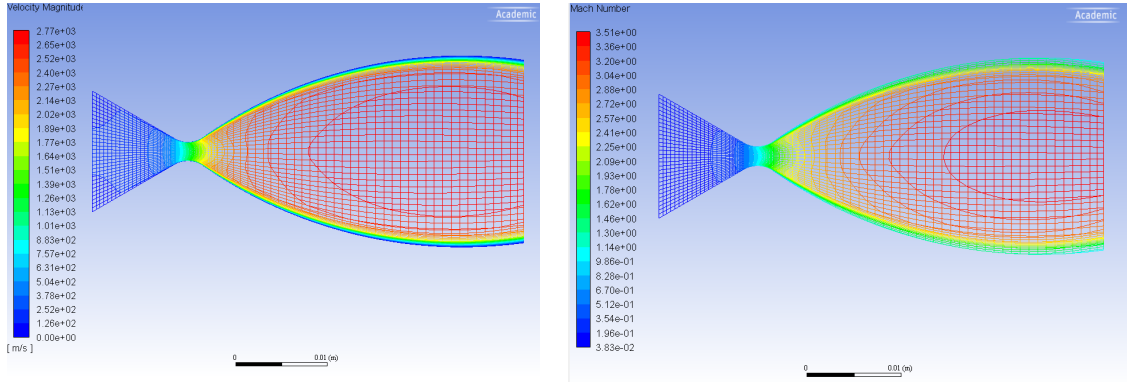

**Figure 7:** ANSYS velocity contour (left) and Mach number contour (right) of the solar-thermal steam propulsion system nozzle.

**Propulsion Performance**

In the previous sections, the performance of the collection system, steam generation system, heating system and the expansion system were analyzed. It is now time to focus on the overall performance of the design. The two variables that characterize the overall performance of a rocket engine are $I_{sp}$ and $\Delta v$.

**Specific Impulse**

Previous section showed the relation between the specific impulse and the steam temperature; and the relation between the concentration ratio and the maximum working temperature. Both concepts can be combined to show the performance of the system in orbit, and the results are shown in Figure 8 (left). Note that the concentration ratio could also represent the direction from the solar irradiance in a concentrator already design at $C = 10,000$.

**Delta-v**

The Tsiolkovsky rocket equation relates the delta-v, this is, the maximum change in velocity of a rocket in absence of external forces, with the effective exhaust velocity and the relation between the initial and final mass of the rocket:

$$\Delta v = v_e \ln \frac{m_0}{m_f} \tag{5}$$

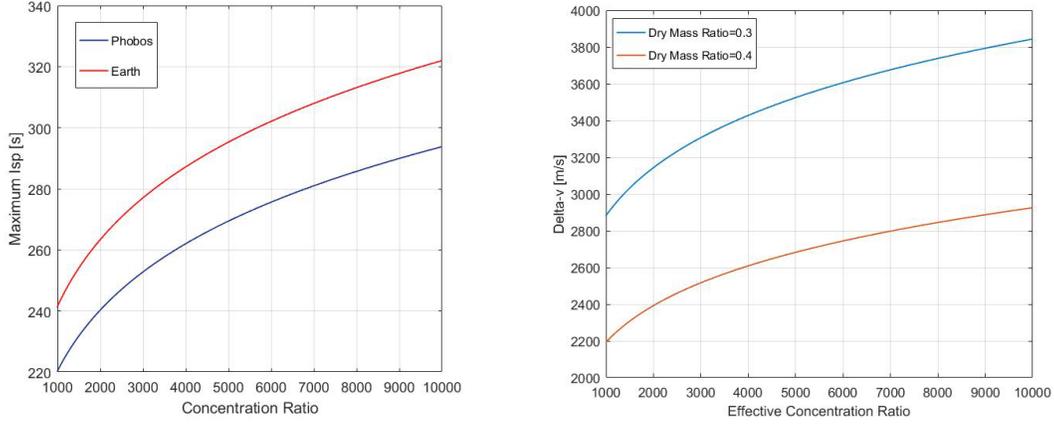

**Figure 8:** (Left) Maximum Isp as a function of the concentration ratio, under Phobos and Earth solar irradiance. (Right) Delta-v as a function of the concentration ratio. Both cases are analyzed under Earth irradiance.

where $v_e$ is the effective exhaust velocity, the logarithm term is the ratio between the total mass of the spacecraft (including the propellant), and the dry mass (without propellant). In Figure 8 (right), the delta-v the system can achieve, as a function of the effective concentration ratio is shown. This effective ratio can also be thought of as the efficiency of a $C = 10,000$ designed system. At the maximum concentration ratio, assuming not misalignment in the solar irradiance and under near-Earth conditions, the spacecraft can achieve a delta-v of 3.8 km/s for a dry mass ratio of 0.3.

Another factor to analyze is the time in which these maneuvers can be performed. This is given by the thrust of the spacecraft in a simple momentum conservation equation. It has been shown that the maximum temperature the propellant can achieve is around 3000K. The total thrust would however also be a function of the mass flow rate of the propellant, that is, how much steam the system is able to heat up in a certain amount of time. Provided that a mass flowrate of about 1g/s can be achieved, this is the concentrator area required as shown in Figure 9 (left). Note, the maximum thrust this system can achieve is 3.2 N. Under this thrust, the spacecraft can perform the 3.8 km/s maneuvering in less than a day and a half. An analysis is performed to understand, under which conditions would this thrust be possible. The standard formula for electrothermal systems can be adapted to photothermal propulsion by:

$$\eta_{opt}\eta_{pt}P = \frac{1}{2}Tv_e \tag{6}$$

where $\eta_{opt}$ is the optical efficiency of the receiver, $\eta_{pt}$ is the solar-to-thermal receiver efficiency and $P$ is the power supply. Optical efficiencies of 0.9 have been previously demonstrated in the industry. In a direct gain concept, the power supply is the product of the solar influx and the concentrator area $P = A_{con}I_{solar}$.

Based on Figure 9, increase in thrust allows the spacecraft to perform this lower delta-v maneuvers in less time, even when the solar-to-thermal energy conversion is lower. This is when lower delta-v maneuvers are required, such as gravity escape or capture orbit from moons are needed, the system can operate at higher thrust and perform the maneuver quickly. When higher requirements such as transit orbits are required, the system would need more time to perform the maneuvering.

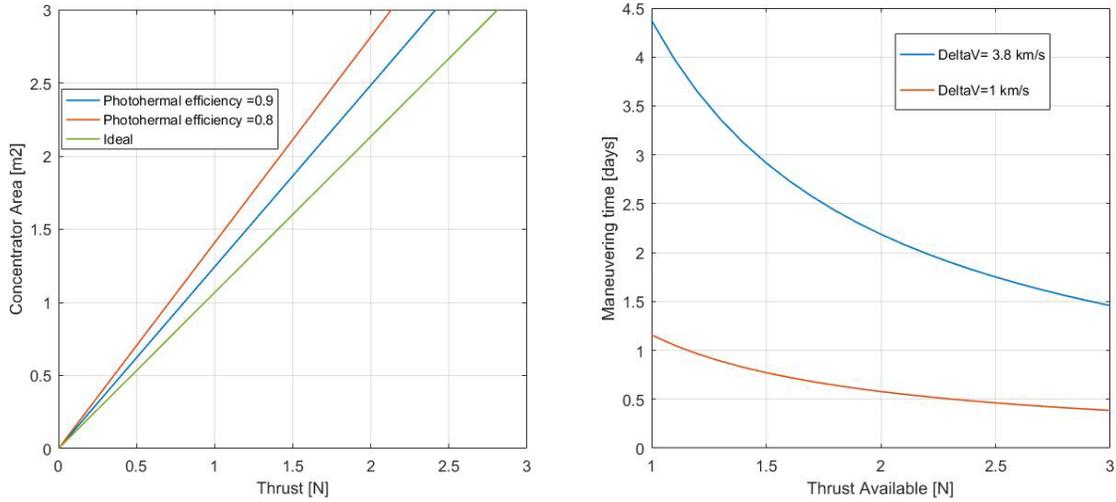

**Figure 9:** (Left) Concentrator Area needed to produce a target thrust. (Right) Maneuvering time as a function of the thrust. The system performance can be tuned for different maneuvers.

## DISCUSSION

In this paper a solar thermal steam propulsion concept was presented. This concept represents a revolutionary solution for interplanetary travel using small satellites. In the past, high specific impulse propulsion systems for these spacecrafts have been developed, such as ion or electromagnetic thrusters. These other solutions however provide low thrust level, which means long maneuvering periods.

Water can be easily extracted from the Moon, C-class Near Earth Objects (NEOs), surface of Mars and Martian Moons Phobos and Deimos and from the surface of icy, rugged terrains of Ocean Worlds. Water is a compelling choice for fuel, it is a green propellant safe and easy to use, it offers modest performance, and can be easily stored for months or years. Using a solar concentrator, heat is used to extract the water which is then condensed as a liquid and stored. Water is then converted into steam and heated using a parabolic dish concentrator. Solar concentration technology has been well studied for terrestrial solar energy applications. In space, rigid structures were demonstrated in the ISUS program. Inflatable structures, with a density below $1\ kg/m^2$ allows truly promising capabilities. In this work, parabolic concentrators optics have been analyzed, and concentrations ratios over 10,000:1 have been found to be possible. Under these concentration ratios, maximum temperatures of 1000 to 3000K have been shown. However, a thermodynamics analysis shows that optical efficiency decreased with the receiver temperature, and therefore the collector performance can be restricted at high temperatures.

Conventional solar collectors capture light energy on an absorbing surface, which is then transferred to the working fluid via conduction or convection. This thermal resistance network can be substantially shorter by using a direct concept. In here, a radiative heat transfer model has been proposed. However, these models are extremely case dependent, and therefore, further experimental work to fully understand the behavior of carbon based nanofluids,

## CONCLUSIONS

In this work, we present a steam-based propulsion that avoids the technological barriers of electrolyzing impure water obtained from off-world sources as propellant. Using a solar concentrator, heat is used to extract the water which is then condensed as a liquid and stored. Steam is then formed using the solar thermal reflectors to concentrate the light into a nanoparticle-water mix. This solar thermal heating (STH) process converts 80 to 99% of the incoming light into heat. The level of thrust achieved will depend in the parabolic dish size, and whether a direct gain or a thermal storage system is employed. As analyzed before, thrust around the Newton level is the maximum achievable for a direct concept and a $1\ m^2$ reflector. In the future, thermal storage technology should be analyzed. Several thermal storage studies have been carried out in the past, but none of them take advantage of nanofluids. The system shows promising capabilities, but its performance will depend on the efficiency of the receiver and the working nanofluid.